\documentclass[runningheads]{svmult}

\usepackage{makeidx}   
\usepackage{graphicx}  
\usepackage{subeqnar}  
\usepackage{multicol}  
\usepackage{physprbb}  
\makeindex             

\begin{document}
\title*{Abundances in Damped Ly$\alpha$ Galaxies}
\toctitle{Abundances in Damped Ly$\alpha$ Galaxies}
%
%
\titlerunning{Abundances in Damped Lyman$\alpha$ Galaxies}
%
\author{Paolo Molaro\inst{1}
}
\authorrunning{Paolo Molaro}
%
%
\institute{Osservatorio Astronomico di Trieste-INAF
   Via G.B. Tiepolo 11,\\
     I-33100 Trieste, Italy}

\maketitle              

\begin{abstract} Damped Ly$\alpha$ galaxies provide a  sample 
of young galaxies  where  chemical abundances can be derived 
throughout the whole universe
with an accuracy
comparable to that for the local universe. Despite  a  large spread in redshift, HI column
density and metallicity,  DLA  galaxies show a remarkable
uniformity in the elemental ratios  rather suggestive of
similar chemical evolution  if not of an unique population. These galaxies
are characterized by  a moderate, if any,  enhancement of  
$\alpha$-elements  over Fe-peak elemental abundance 
with [S/Zn]$\approx$ 0 and [O/Zn]$\approx$ 0.2, rather similarly to the dwarfs
galaxies in the Local Group.
Nitrogen  shows   a peculiar behaviour  with a 
bimodal distribution and possibly two plateaux. In particular, the
 plateau at 
    low N   abundances 
([N/H] $<$ -3),  is not observed in other atrophysical sites and might be 
evidence  for primary N production by massive  stars.   
\index{abstract} 
\end{abstract}

\section{Introduction}
Any slab of   intervening material along the 
line of sight of a background source with hydrogen column density   high 
enough to produce  damping wings in Ly$\alpha$, conventionally $\log N(HI)$ $>$ 20.3 $cm^{-2}$, is 
producing a Damped Ly$\alpha$ galaxy (DLA). 
Damped  systems  hold a large fraction of neutral
 gas at high redshift and are 
  considered the progenitors of present day galaxies.  
  Holding   neutral gas 
and being free  from
ionization effects, DLAs  provide  column densities   at  very high 
precision ($\approx$ 10 \%), which  together with the fact that 
they    are observed up to  redshift $\approx$ 4.5 or equivalently  
  at  a look-back time of $\approx$ 12   Gyr, 
  expand to almost the entire  universe the possibility of a 
 detailed chemical investigation. 
 Observed metallicities  are generally low, with
 [Fe/H] varying between -2.5 and -1.0, but never below 
  -2.5 and with a  mild evolution with redshift. 
  The elemental abundances, which we discuss here in more detail,  resemble very closely  those
  observed in the dwarf and irregular galaxies of the Local Group which are 
  the subject of this conference.

 \section{Dust}

Dust is probably present in the DLA and  significantly affects 
  the observed abundances. I counted 
  55 systems for which both Fe 
and Zn are measured,  which are   plotted in Fig. 1. 
The abundance of Fe is always found below that of Zn. 
 By analogy with the interstellar 
medium, this behaviour is interpreted as the effect of 
 some Fe being  locked into dust grains. 
 Other indicators for the presence of dust are  the correlation of [Fe/Zn]   with H$_{2}$  and the reddening excess 
of   QSOs behind Damped galaxies.
The 
figure shows a clear  trend of Fe depletion with  metallicity. 
There is also evidence for a sort of  
threshold line [Fe/Zn] $\approx$ -(2 +[Zn/H])
with  no Fe  depletion below it, which   points out to the presence of 
a forbidden region 
  for   dust formation.  
This behaviour may be relevant for the  understanding of  
dust formation  in low metallicity 
environments.

\begin{figure}[h]
\begin{center}
\includegraphics[width=.45\textwidth,angle=-90]{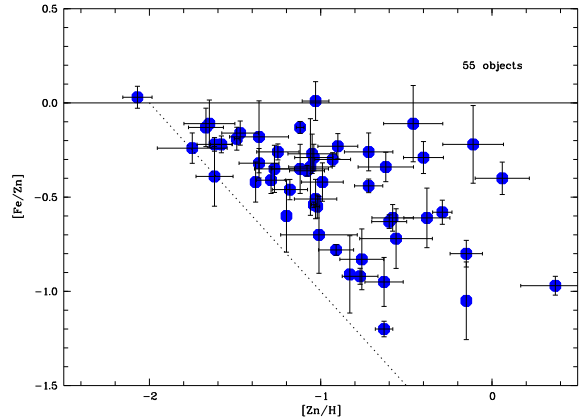}
\end{center}
\caption[]{Dust evidence in the DLAs}
\label{eps1}
\end{figure}

\section{The [$\alpha$/Zn] ratio}

The presence of dust affects   abundances of 
elements participating  into dust grains, and it 
 needs either to be corrected or to be avoided. 
 Considering the uncertainties involved in  dust 
formation, the second approach looks safer. 
  S and O are $\alpha$
elements  showing little 
affinity with dust, while   
Zn is, rather uncomfortably, the only one available  
among the iron-peak group. 
In Fig 2   the [S/Zn] elements for 24 
Damped galaxies are shown. The average value is
 consistent with zero: $<[S/Zn]>$ = 0.01 $\pm$ 0.14. 
There is also a hint of 
  decreasing [S/Zn] with the increasing of 
metallicity. 
O has been   measured in a limited number of objects and those in which also
 Zn is available 
 are  plotted in Fig 3, together with a couple of particularly low [O/Fe], 
 which are a significant 
upper bound  to the [O/Zn] value.  
QSO 0347-3819 is generally  
considered  the best case for   
evidence of $\alpha$ enhancement.  However, 
the figure shows   a  new [O/Zn] measure from
 a revised measurement of ZnII 
 by Levshakov et al ($\log N(ZnII) = 12.26$ cm
$^{-2}$,  2004, private com.). 
 The revised ratio is  [O/Zn]=0.1$\pm$0.1 and there 
is no more  strong  $\alpha$ enhancement  in this DLA. 
The case of 
Q 0812+32, with [O/Zn]=0.35,  should be probably considered separately 
since this DLA is rather peculiar and is
considered the  progenitor 
of an elliptical. 
Thus, with the exception of the QSO 0812+32 case,  
all the DLA  galaxies show  mild or absent  
$\alpha$ enhancement, reminischent of what found in the stars of the 
Local Group galaxies.
To increase the statistical significance of the result  one can use
 Si that, despite being  refractory, 
tracks S rather well on average. In 40 DLA  galaxies [Si/Zn] $\approx$ 0,  with only few exceptions of 
particularly dusty objects, showing  that the absence of $\alpha$ enhancement  is 
   not due to poor statistics but  is a rather general property as was first
   suggested by Molaro et al (1996)

\begin{figure}[h]
\begin{minipage}[t]{0.5\linewidth} 
\begin{center}
\includegraphics[width=.65\textwidth,angle=-90]{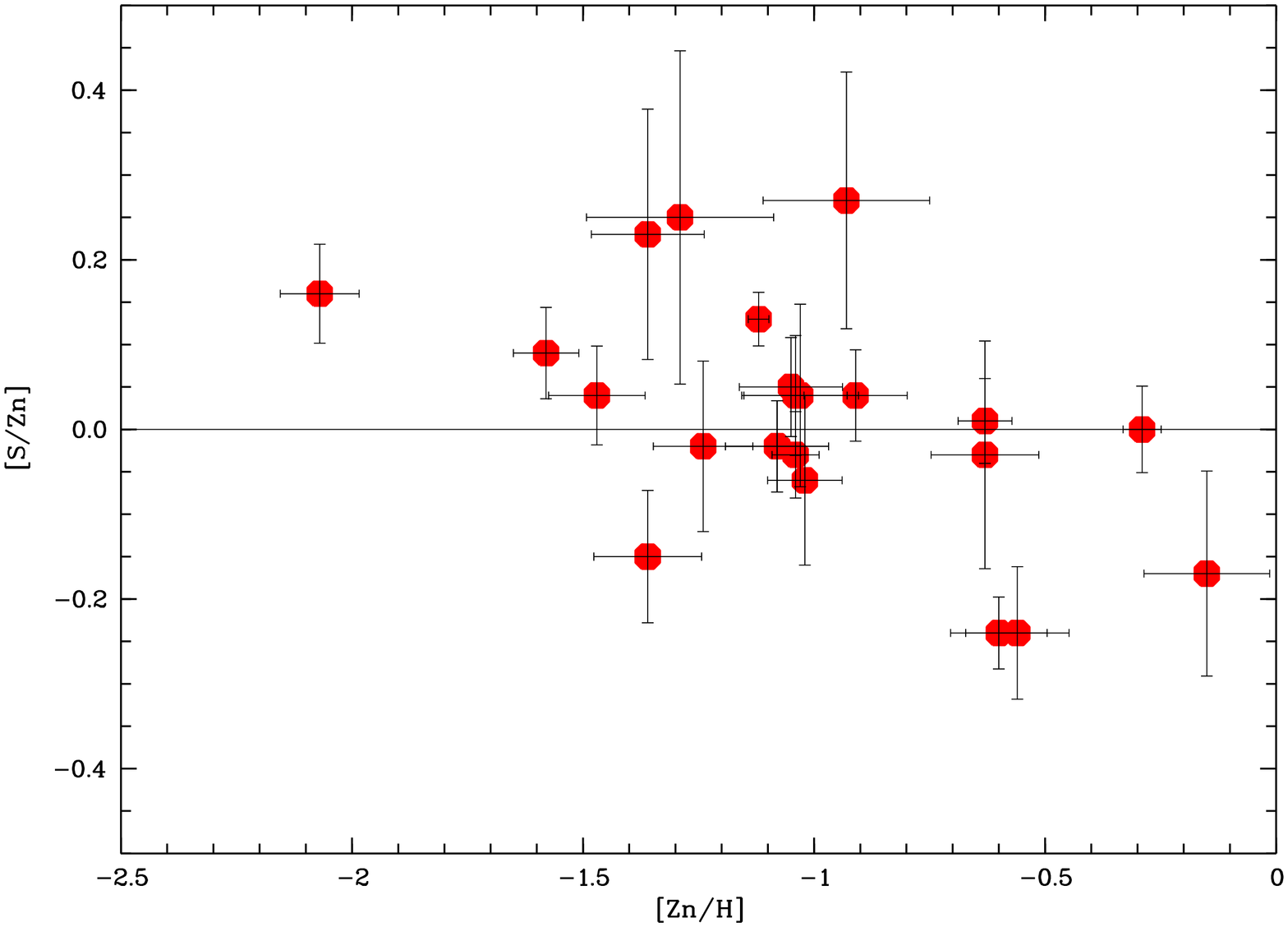}
\end{center}
\caption[]{[S/Zn]  in DLAs}
\label{eps2}
\end{minipage}
\begin{minipage}[t]{0.5\linewidth}
\begin{center}
\includegraphics[width=.65\textwidth,angle=-90]{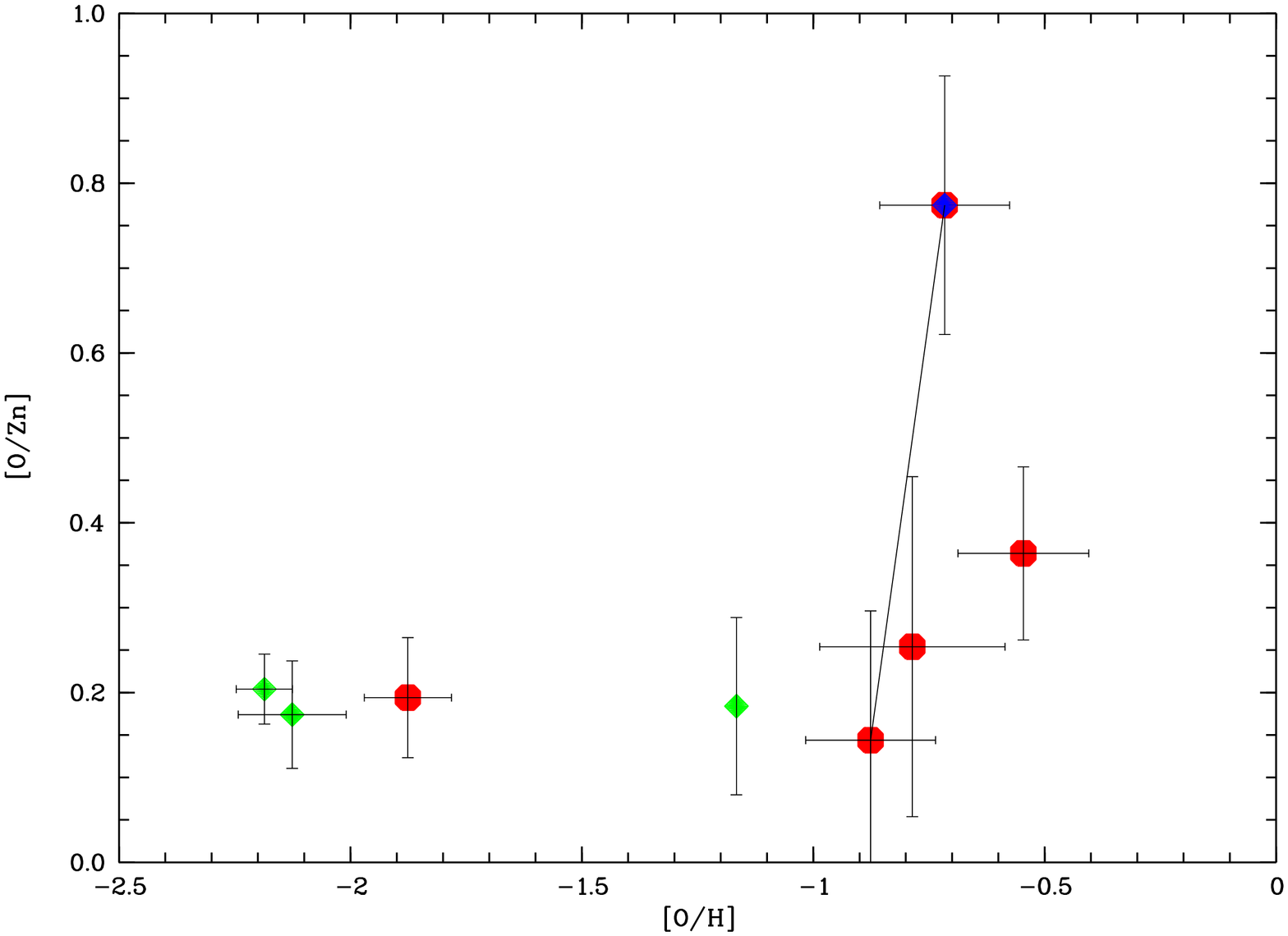}
\end{center}
\caption[]{[O/Zn]  in DLAs, with some [O/Fe]}
\label{eps3}
\end{minipage}
\end{figure}

\subsection{Nitrogen}

N shows little affinity with dust and is thought to be produced by intermediate mass 
stars in the 4-8 M$_{0}$ range which undergo HBB 
 in the AGB  phase.
N  has a 
dual nature since it is primary at low metallicities 
and secondary at high metallicities,  as  
observed in the extragalactic HII regions. 
Being 
produced by relatively small masses it is  produced with
a time delay compared to O or other $\alpha$ a fact that can explain 
    the  rather scattered [N/Si] at low abundances. It has 
been suggested by Prochaska et al (2002) that the 
DLAs show a bimodal distribution rather than 
a pure scatter behaviour. From the 2 systems 
initially considered by them  there are now 5 systems and 
2 upper limits that support the proposal for a bimodal distribution. 
The existence of two  different regimes   can be   better appreciated when  [N/Si] 
is plotted versus [N/H] as shown in Fig. 4. 
The plot  clearly shows that the 
in DLA low and high [N/Si] occupy two  separate regions. A possible 
explanation  is that the low values  are young systems where N comes from massive stars and the AGB 
have not yet started to produce the bulk of  N.    
A  primary N production by massive stars is not foreseen by standard 
models but it is present in 
the zero metallicity  
  models of Chieffi \& Limongi  (2002)
and the predicted [N/Si] ratios after integration over an 
appropiate IMF show ratios  
close to the observed ones. It would be rather appealing  to find  this N as 
{\it the smoking gun} of zero metal stars. 
In Fig 4 the   [N/Si] of DLAs are  shown together with the recent N 
determinations provided  in halo stars. 
Stellar values are in rough agreement with this picture,  showing no values 
 below [N/Si]=-1.5, although
they populate also the portion of the diagram with [N/H] $<$ -3.0.

\begin{figure}[h]
\begin{center}
\includegraphics[width=.46\textwidth,angle=-90]{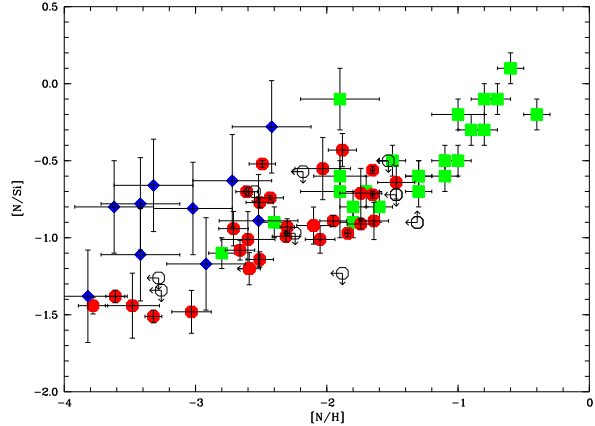}
\end{center}
\caption[]{Nitrogen in DLAs, circles, and halo stars  from Spite et al (2004),
losange, and from Israelian et al
(2004), squares.} 
\label{eps4}
\end{figure}

\subsection{A single population?}

 Determination of  relative abundances of elements produced differently  by Type I, TypeII SNae and 
 IMS  probes the early chemical build-up. 
 Chemical evolutionary models characterized by a
 relatively low star formation rate have been proposed  to explain the mild [$\alpha$/Zn]
  and the  [N/Si] ratios in the DLA. A low SFR is commonly found in  dwarfs, 
  irregulars,  LSB galaxies 
  or in 
  the external part of rotational disks. 
  Observationally the DLAs show a remarkable constancy in the observed
 elemental ratios.  So far there are  15 DLAs  for which we have 
  the whole  triplet of elements N, Si and Zn. Si in this case is
 used as a fair proxy for O or S.
 The observed elemental ratios are:
  $<[S/Zn]>$ = 0.01 $\pm$ 0.18 and $<[N/Si]>$ = -0.79 $\pm$ 0.16, i.e.
  remarkably  uniform considering that  the metallicity in this sample
   is varying from -2.0 to -0.5, redshift  from 1.6 to 3.4 and  N(HI)
  from 20.0 to 21.5. Such a homogeneitiy in the ratios is strongly suggestive of  
  very similar evolutionary properties of the
  DLAs if not of a specific  population of galaxies, which might dominate the
  DLA class at least at high redshift.
  
%

%

\end{document}